\documentclass[preprint,aps,showpacs,floatfix]{revtex4}
\usepackage{epsfig}
\usepackage{bm}

\begin{document}

\title{
Lorentz boosted nucleon-nucleon potential  
applied to the
\bm{$ \vec{^3{\rm He}} ({\vec e},e' p)pn$}
and
\bm{$ \vec{^3{\rm He}} ({\vec e},e' n)pp$}
processes
}

\author{J.~Golak, R.~Skibi\'nski, H.~Wita{\l}a}
\affiliation{M. Smoluchowski Institute of Physics, Jagiellonian University,
                    PL-30059 Krak\'ow, Poland}

\author{W.~Gl\"ockle}
\affiliation{Institut f\"ur Theoretische Physik II,
         Ruhr Universit\"at Bochum, D-44780 Bochum, Germany}

\author{A.~Nogga}
\affiliation{Forschungszentrum J\"ulich, IKP
(Theorie), D-52425 J\"ulich, Germany}

\author{H.~Kamada}
\affiliation{Department of Physics, Faculty of Engineering,
  Kyushu Institute of Technology,
  1-1 Sensuicho, Tobata, Kitakyushu 804-8550, Japan}

\date{\today}

\begin{abstract}
We formulate an approximate relativistic framework
for an analysis of the 
$ \vec{^3{\rm He}} ({\vec e},e' p)pn$
and
$ \vec{^3{\rm He}} ({\vec e},e' n)pp$
reactions.
 Restricting the rescattering series to one term linear in the two-nucleon (2N)  
t-matrix we incorporate various relativistic features when calculating 
a nuclear current matrix element. These relativistic ingredients encompass  
 the relativistic $^3$He wave function
based on the concept of the Lorentz boosted nucleon-nucleon potential 
together with the boosted 2N t-matrix, relativistic kinematics 
 and relativistic single-nucleon current operator. 
 This allows us to estimate the magnitude of relativistic effects
not included in the standard nonrelativistic approach.
\end{abstract}
\pacs{21.45+v,21.10-k,25.10+s,25.20-x}
\maketitle



\section{Introduction}
\label{sec:1}

Modern three-body calculations allow for a quantitative
description of the three-nucleon (3N) system not only in the bound state
\cite{nogga03}
but also for the continuum states (see for example \cite{report96,kuros02}). 
This gives the
possibility to test our understanding of the three-body system via interactions
with external probes. Among many processes which can be listed here,
electron scattering on $^3$He is of special importance \cite{report,deltuva04}.
This process serves as a rich source of information 
about the nucleon form factors \cite{xu00,extraction,sensitivity,xu03}
and important properties of the $^3$He 
nucleus \cite{correlations,spindep,polarizations}.

Electron induced breakup of $^3$He involves many components of the dynamical 
scenario. 
  Among them the initial $^3$He and final scattering
states must be  calculated consistently for the same 3N Hamiltonian comprising 
not only two-nucleon (2N) but also three-nucleon (3N) forces.
Consequently, also many-body currents consistent with those forces 
should be taken into account.
We refer the reader to \cite{report} for a detailed  discussion
of the numerical techniques necessary to perform calculations
 of this reaction. 
Currently  this can be done only nonrelativistically, which is a major restriction and 
leads to   serious difficulties in interpretation of 
 many experiments performed at high energy and momentum transfers.
 Due to large differences between the nonrelativistic 
and relativistic kinematics an analysis of such experiments can not be 
 undertaken within a strictly nonrelativistic framework.

We are not aware of any consistent, relativistic 3N scattering calculation.
Also in the present paper we report about a less rigorous approach to
the description of the $ \vec{^3{\rm He}} ({\vec e},e' p)pn$
and $ \vec{^3{\rm He}} ({\vec e},e' n)pp$ processes.
This approach does not include all final state interactions (FSI)
among the three outgoing nucleons but restricts the rescattering 
to only one ``spectator'' pair of nucleons which is assumed not to take part
in the photon absorption. There are definitely kinematical regions where such a
reaction mechanism seems to be plausible. Furthermore this approximation was 
used successfully in the analysis of many experiments (see for example
\cite{xu03,carasco03}).

We would like to add to this treatment of 
electron induced breakup of $^3$He new truly relativistic features.
We continue work started in \cite{kamada02}, where first steps
to extend the Hamiltonian scheme in equal time formulation to 3N scattering
were made. To this aim the Lorentz boosted nucleon-nucleon (NN) 
potential which generates
the NN $T$-matrix in a moving frame via a standard Lippmann-Schwinger equation
was calculated and applied to the 3N bound state problem.
In the present paper we show how to obtain the (antisymmetric) 
 3N relativistic wave function and formulate an approximate framework 
which can be used as a practical tool for an analysis of experimental results,
for example in quasi-elastic reactions at high energy and momentum transfers.

We give the reader a detailed derivation of our formalism in Sec.~\ref{sec:2}.
Section~\ref{sec:3} shows our results for the semi-exclusive three-body
breakup of $^3$He. We end with a brief summary in Sec.~\ref{sec:4}.

\section{Theory}
\label{sec:2}

Before we remind the reader of the most important ideas 
about the Lorentz boosted NN potential,
it seems appropriate to start with the well known nonrelativistic 
concepts.
 
The nonrelativistic 2N bound state $ \mid \psi^{(nr)}_b \rangle $
obeys the equation
\begin{eqnarray}
\mid \psi^{(nr)}_b \rangle = G_0^{(nr)} \, v^{(nr)} \, 
\mid \psi^{(nr)}_b \rangle ,
\label{first}
\end{eqnarray}
where $ v^{(nr)} $ is the nonrelativistic NN potential and
$G_0^{(nr)}$ is the nonrelativistic 2N free propagator. 
This can be written in the 2N center of mass (c.m.) frame 
 by projecting onto  the  eigenstate of relative 
momentum $\vert{\vec p}>$ (${\vec p}$ and $-{\vec p}$ are then the individual 
 nucleon momenta)
\begin{eqnarray}
\psi^{(nr)}_b ( {\vec p}\, ) = {1 \over { M_b - 2 m -  
\frac { {\vec p}^{\, 2} }{m}  } } \,
\int d^3 p' \, v^{(nr)}( {\vec p},{\vec p}^{\, \prime}) \, 
\psi^{(nr)}_b ( {\vec p}^{\, \prime} ) .
\label{psinrl}
\end{eqnarray}
Here $M_b$ is the 2N bound state rest mass and $m$ is the nucleon mass. 
 The corresponding Lippmann-Schwinger equation 
\begin{eqnarray}
t^{(nr)}  \, = \,  v^{(nr)} \, + \, t^{(nr)} \, G_0^{(nr)} \, v^{(nr)}
\end{eqnarray}
for the 
$t$-matrix $t^{(nr)}$ takes in the momentum space the following form
\begin{eqnarray}
t^{(nr)}({\vec p}  , {\vec p}^{\, \prime} \, ) 
\, = \, v^{(nr)}({\vec p} , {\vec p}^{\, \prime} \,) \, + \,
\int \! \! d^3 p^{\, \prime\prime} \  {{
t^{(nr)}( {\vec p} , {\vec p}^{\, \prime\prime})
v^{(nr)}( {\vec p}^{\, \prime\prime} , {\vec p}^{\, \prime} \, ) }
\over {  
E_{12}^{nr} - \frac{{\vec p}^{\, \prime \prime \, 2}}{m} 
+ i\epsilon }} ,
\label{R-LS0}
\end{eqnarray}
where $E_{12}^{nr} $ is the nonrelativistic 2N c.m. kinetic energy. 
The Galilean invariance of the nonrelativistic scenario guarantees that the 
relative momentum and Eqs.(\ref{first})-(\ref{R-LS0}) remain 
 frame independent. 

On the other hand, a relativistic NN potential $v^{(rl)}$ defined in the 
2N c.m. system appears in the relativistic bound state equation 
\begin{eqnarray}
\mid \psi^{(rl)}_b \rangle = G_0^{(rl)} \, v^{(rl)} \, 
\mid \psi^{(rl)}_b \rangle ,
\label{psirl}
\end{eqnarray}
and in the relativistic form of the Lippmann-Schwinger equation
\begin{eqnarray}
t^{(rl)}  \, = \,  v^{(rl)} \, + \, t^{(rl)} \, G_0^{(rl)} \, v^{(rl)} .
\label{tmrl}
\end{eqnarray}
In the momentum space  spanned by eigenstates of the 2N c.m. relative 
momentum ${\vec p}$ the Eqs.~(\ref{psirl}) and (\ref{tmrl}) can be written as
\begin{eqnarray}
\psi_b^{(rl)} ( {\vec p} ) = {1 \over { M_b - \omega ( {\vec p}\, ) } } \,
\int d^3 p' \, v^{(rl)} ( {\vec p},{\vec p}^{\, \prime}) \, 
\psi_b^{(rl)} ( {\vec p}^{\, \prime}\, )
\label{psirlmom}
\end{eqnarray}
and
\begin{eqnarray}
t^{(rl)} ({\vec p} , {\vec p}^{\, \prime}\, ) = v^{(rl)} 
({\vec p} , {\vec p}^{\, \prime}\, ) + 
\int d^3 p^{\, \prime\prime}\, \ {{
t^{(rl)} ( {\vec p} , {\vec p}^{\, \prime \prime} \,)
v( {\vec p}^{\ \prime\prime \,} , {\vec p}^{\, \prime}\, ) }
\over { E_{12}^{rl} - \omega ( {\vec p}^{\, \prime}) + i\epsilon }} ,
\label{tmrelmom}
\end{eqnarray}
where
\begin{eqnarray}
\omega ( {\vec p}) \equiv 2 \sqrt{ m^2 + {\vec p}^{\, 2} }
\label{omegadef}
\end{eqnarray}
and the relativistic 2N c.m.  energy 
is denoted by $E_{12}^{rl}$.  

A Lorentz boosted nucleon-nucleon potential $ V ( \vec q \, )$ was introduced 
in \cite{gloeckle86} in order to generalize the concept of the relativistic
potential for 2N systems with the non-zero total momentum $\vec q$.
It is formally defined via 
\begin{eqnarray}
V  ( \vec q \, )  \,  \equiv \, \sqrt{ \left[ \omega ( {\vec p} \, ) + v^{(rl)}  \right]^2 \, + \,
{\vec q}^{\ 2} }  \ - \ 
\sqrt{ ( \omega ( {\vec p}\, ) )^2 \, + \, {\vec q}^{\ 2} }
\end{eqnarray}
and by construction fulfills $V ( \vec q \, = 0  ) \,  = \, v^{(rl)} $.
The non-trivial task of obtaining matrix elements 
$  V(  {\vec p} ,  {\vec p}^{\ \prime} ; \vec q \, )$ for arbitrary $ \vec q $
was accomplished in \cite{kamada02}. 

With use of the boosted potential, the equation for the relativistic 2N bound 
state moving with the total momentum $\vec q$ reads
\begin{eqnarray}
\psi_b^{(rl)} ( {\vec p} \, ) = 
{1 \over { \sqrt{M_b^2 + {\vec q}^{\  2} } 
 - \sqrt{ \omega ( {\vec p} \, )^{\, 2} + {\vec q}^{\ 2} }  } } 
\int d^3 p' \, V( {\vec p} , {\vec p}^{\, \prime} \, ; \, {\vec q}\, ) \, 
\psi_b^{(rl)} ( {\vec p}^{\ \prime} ),
\label{psibst}
\end{eqnarray}
so the boosted potential allows us to preserve 
the same structure of the equation as
in (\ref{psinrl}) and (\ref{psirlmom}). Note 
that $ \psi_b^{(rl)} ( {\vec p} ) $
appearing in Eqs.~(\ref{psirlmom}) and (\ref{psibst}) are identical, i.e. 
the wave function is represented in a way, which does not depend 
on ${\vec q}$. 
This is possible because the relative momenta $\vec p$ 
and $ {\vec p}^{\ \prime} $
in both cases are defined in the 2N c.m. system.

A formalism for treating the relativistic three-body Faddeev equations was
introduced in \cite{gloeckle86.2,gloeckle86}. 
Since the formal structure of the 3N Hamiltonian
\begin{eqnarray}
 H = H_0 + \sum\limits_{i < j} V_{ij}, 
\label{Hrel}
\end{eqnarray}
is the same for relativistic and nonrelativistic approaches, the formal 
derivation of the Faddeev equations is also the same 
in both cases \cite{gloeckle86}.
(Note that $V_{ij}$  in \ref{Hrel} is the boosted 2N potential.)
Thus the Faddeev component $\mid \Phi \rangle $
of the 3N relativistic wave function $ \mid \Psi \rangle $ 
generated by interaction $V$ in the 2N subsystem obeys
\begin{eqnarray}
\mid \Phi \rangle \ = \ G_0 \, T \, P \mid \Phi \rangle ,
\label{Phi}
\end{eqnarray}
where $T$ is the Lorentz boosted $T$-matrix generated by potential $V$, 
$G_0$ is the relativistic 3N free propagator
and $P$ is a permutation operator which accounts for 
 the fact that we treat nucleons as identical particles.
It is given in terms of the transposition $P_{ij}$ interchanging 
nucleons ``i'' with ``j'':  
$P \equiv P_{12} P_{23} + P_{13}P_{23}$.
The wave function $\mid \Psi \rangle $  follows from the Faddeev 
component via 
\begin{eqnarray}
\mid \Psi \rangle =  (1 + P) \mid \Phi \rangle .
\label{Psi}
\end{eqnarray}

In \cite{gloeckle86} the boosted $T$-matrix  
is constructed from the 
relativistic 2N $t$-matrices of Eq.(\ref{tmrelmom})
in a quite complicated way.
Since we have now the Lorentz boosted potential $V ( {\vec q}\, )$ at 
our disposal,
we can obtain the boosted (off-shell) $T$-matrix directly 
via the Lippmann-Schwinger equation which, when written in the 3N c.m. 
system, takes the form 
\begin{eqnarray}
T( {\vec p} , {\vec p}^{\, \prime}  \, ; \, {\vec q} \,) =
V( {\vec p} , {\vec p}^{\, \prime}  \, ; \, {\vec q} \,)
\ + \
\int d^3 p^{\, \prime\, \prime} {{
T({\vec p} , {\vec p}^{\, \prime\prime} \, ; \, {\vec q} \, ) \, 
V( {\vec p}^{\, \prime\, \prime} , {\vec p}^{\, \prime} \, ; \, {\vec q} \, ) }
\over { 
E_{3N} - \sqrt{m^2 + q^2}  
\, - \,
\sqrt{ \omega ( {\vec p}^{\, \prime\prime} )^{\, 2} + {\vec q}^{\ 2} } 
+ i\epsilon }} ,
\label{tbst}
\end{eqnarray}
where
$E_{3N}$ is the total energy of the 3N system and $\vec q$ is the momentum of the 
spectator nucleon ($-\vec q$ is then a total momentum of the 2N subsystem). 
Due to the following observation, Eq.~(\ref{tbst}) can be 
solved as easily as Eq.~(\ref{R-LS0}).
Namely defining 
\begin{eqnarray}
f_q ( p ) \equiv 
\sqrt{ \frac{E_{3N} - \sqrt{m^2 + q^2}   
\ + \ \sqrt{4 m^2 + 4 p^2 +q^2 }}{ 4 \, m}   } ,
\end{eqnarray}
\begin{eqnarray}
v( {\vec p} , {\vec p}^{\, \prime} \, ; \, {\vec q}) \equiv 
f_q ( p ) \, V( {\vec p} , {\vec p}^{\, \prime} ; \, {\vec q}) \, 
f_q ( p^\prime \, ) ,
\end{eqnarray}
\begin{eqnarray}
t( {\vec p} , {\vec p}^{\, \prime} \, ; \, {\vec q}) \equiv 
f_q ( p ) \, T( {\vec p} , {\vec p}^{\, \prime} \, ; \, {\vec q}) \, 
f_q ( p^\prime \, )
\label{taux}
\end{eqnarray}
and
\begin{eqnarray}
p_0^2  \equiv 
\frac14 \, \left( (E_{3N} - \sqrt{m^2 + q^2}  )^2 - 4 m^2 - q^2 \right) 
\end{eqnarray}
we arrive at
\begin{eqnarray}
t( {\vec p} , {\vec p}^{\, \prime}  \, ; \, {\vec q} \,) =
v( {\vec p} , {\vec p}^{\, \prime}  \, ; \, {\vec q} \,)
\ + \
\int d^3 p^{\, \prime\prime} \, {{
t({\vec p} , {\vec p}^{\, \prime\prime} \, ; \, {\vec q} \, ) \, 
v( {\vec p}^{\, \prime\, \prime} , {\vec p}^{\, \prime} \, ; \, {\vec q} \, ) }
\over {
\frac{p_0^2}{m}  - \frac{{\vec p}^{\, \prime \prime \, 2}}{m} 
+ i\epsilon 
}} .
\label{R-LS02}
\end{eqnarray}
This looks like a nonrelativistic Lippmann-Schwinger equation (\ref{R-LS0})
and can be solved by the same techniques. Once this equation is solved,
Eq.~(\ref{taux}) is used to get the 
$T( {\vec p} , {\vec p}^{\, \prime}  \, ; \, {\vec q} \,)$ matrix elements.
Note that 
$ p_0^2$ in (\ref{R-LS02}) might be in a general case also negative.

As shown in \cite{kamada02} one needs matrix elements of the relativistic 
potential $v^{(rl)}$ in order to obtain  
$V( {\vec p}^{\, \prime} , {\vec p} \, ; \, {\vec q} ) $. 
The boosted potential is then given by the NN bound state wave function 
and the half-shell NN t-matrices obtained in the 2N c.m. system. The only 
requirement on $v^{(rl)}$ is that it should describe properly existing 2N data
 set. 
 It is possible to construct $v^{(rl)}$ directly (see for example 
\cite{carlson93}) or start 
with a particular modern nonrelativistic potential $v^{(nr)}$ and apply 
  a scale transformation from \cite{kamada98} to generate 
a phase equivalent relativistic
potential $v^{(rl)}$.  This second method was criticized in \cite{allen00} but 
nevertheless it remains a possibility for practical calculations.
 Since the general expression for boosted potential 
$ V ( {\vec p}, {\vec p}^{\, \prime} \, ; \, \vec q ) $
given in \cite{kamada02} is quite complicated  it is desirable
to find an approximation simplifying numerical calculations.
A simple choice is a restriction to the leading order term in a 
$q/\omega$ and $v/\omega$ expansion 
\begin{eqnarray}
 V ( {\vec p}, {\vec p}^{\, \prime} \, ; \, \vec q ) \approx 
 v^{(rl)} ( {\vec p}, {\vec p}^{\, \prime} ) \left(
1 - \frac{ {\vec q}^{\, 2} }{2 \, \omega( {\vec p} \, ) \, 
\omega( {\vec p}^{\, \prime} \, ) } 
\right) ,
\label{approxboost}
\end{eqnarray}
what turned out to be sufficient for a wide range 
of $\mid \vec q \mid $ values \cite{witala05} (see also \cite{wallace01}).
Such an approximation results in a moving deuteron wave function, a binding 
energy and S- and D-state  probabilities very close to the ones for the 
deuteron at rest. 

In a recent paper \cite{keister06} an alternative way to arrive directly
at the boosted NN $t$-matrix is given. This is without approximation
and appears easy to be implemented. Unfortunately, we got aware of that paper 
only after finishing this study.

We have now all ingredients to write Eq.~(\ref{Phi}) in the momentum space.
It reads in the 3N c.m. system \cite{gloeckle86,gloeckle86.2} 
\begin{eqnarray}
\Phi(\vec p,\vec q)=  { 1 \over E_b - {\cal E } (\vec p , \vec q) }  
\int d ^3 q^{\, \prime} \, { T_a ( \vec p , \vec \pi ( {\vec q}^{\ \prime} , 
- \vec q- {\vec q}^{\ \prime}); {\vec q} \, )
\over {{\cal N }  ( {\vec q}^{\, \prime} , -{\vec q} 
- {\vec q}^{\, \prime} \, ) \,
{\cal N } ( -{\vec q} - {\vec q}^{\, \prime} , {\vec q} \, ) }} \;
\Phi(\vec \pi (-{\vec q} - {\vec q}^{\, \prime} 
, {\vec q} \, ), {\vec q}^{\, \prime} \, ) 
\label{Faddeevrelmom}
\end{eqnarray}
where $E_b$ is the 3N binding energy and the index $``a''$ in the 
boosted $T$-matrix 
indicates that it is the properly antisymmetrized operator with respect 
to exchanges of two interacting nucleons.  
The vector $\vec p$ represents the
relative momentum of two interacting nucleons in their 2N c.m. subsystem, 
 and $\vec q$ stands for the
momentum of the spectator nucleon ($-\vec q$ is the total momentum 
of the interacting 2N subsystem). 
 The kinetic energy ${\cal E }$ is given by
\begin{eqnarray}
{\cal E }( \vec p , \vec q ) = 
\sqrt{\omega \left( \vec p \, \right)^2  + {\vec q}^{\, 2} } 
+ \sqrt{ m^2 +  {\vec q}^{\, 2} } - 3 m  .
\label{kinetic}
\end{eqnarray}
Let us denote the individual momenta of the three nucleons in their  
3N c.m. system by
${\vec p}_i$, their total energies 
by $E_i$, and assume that nucleon 1 is the spectator.  
Then the relation between the  momenta ${\vec p}_2$ and $ {\vec p}_3$ 
 of two interacting nucleons and their 2N c.m. 
 relative momentum ${\vec p}$, and between spectator momentum  ${\vec q}$ and 
 the total  momentum $ {\vec p}_{23}$ of the interacting 2N subsystem  reads
\begin{eqnarray}
\vec q = {\vec p}_1 = -  \left( {\vec p}_2  + {\vec p}_3 \, \right) \equiv 
- {\vec p}_{23} ,
\label{p23total}
\end{eqnarray}
\begin{eqnarray}
\vec p \equiv {\vec \pi} \, ({\vec p}_2, {\vec p}_3) \equiv  { 1 \over 2 } \,
\left(  {\vec p}_2  - {\vec p}_3 \right)
- { 1 \over 2 } \, {\vec p}_{23}
\, \left[
 {  E_2  - E_3 \over ( E_2 + E_3 ) +
\sqrt{ (E_2 + E_3)^2 - 
  {\vec p}_{23}^{\ 2} 
  } } \right] .
\label{p23relative}
\end{eqnarray}
Relations (\ref{p23total}) and (\ref{p23relative}) can be inverted
to express the individual momenta ${\vec p}_2 $, ${\vec p}_3$ in the 
3N c.m. system 
in terms of the relative momentum ${\vec p}$ of the (23) pair in its
2N c.m. system and its total momentum ${\vec p}_{23}$ in the 
3N c.m. system: 
\begin{eqnarray}
{\vec p}_2 =  {\vec p} + \frac12{\vec p}_{23} \, 
+ \frac{{\vec p} \cdot {\vec p}_{23}}
{ \left[ 
\omega \left( \vec p \, \right) + \sqrt{ \omega \left( \vec p \, \right)^2 
+ {\vec p}^{\ 2}_{23} }
\right]
 \omega \left( \vec p \, \right)  } \, {\vec p}_{23} ,
\label{p2}
\end{eqnarray}
\begin{eqnarray}
{\vec p}_3 = -{\vec p} + \frac12{\vec p}_{23} \, 
- \frac{{\vec p} \cdot {\vec p}_{23}}
{ \left[ 
\omega \left( \vec p \, \right) + \sqrt{ \omega \left( \vec p \, \right)^2 
+ {\vec p}^{\ 2}_{23} }
\right]
 \omega \left( \vec p \, \right)  } \, {\vec p}_{23} .
\label{p3}
\end{eqnarray}

The two additional
factors
$ {\cal N }  ( {\vec q}^{\, \prime} , -{\vec q} - {\vec q}^{\, \prime} \, )$
and ${\cal N } ( -{\vec q} - {\vec q}^{\, \prime} , {\vec q} \, )$
in Eq.~(\ref{Faddeevrelmom}) 
which generally can be written as \cite{gloeckle86}
\begin{eqnarray}
{\cal N }( {\vec p}_2, {\vec p}_3) = \left\vert {\partial ({\vec p}_2
 , {\vec p}_3 ) \over \partial ( {\vec p}, {\vec p}_{23} )}
 \right \vert ^{\frac12}  
=
 \left( 4E_2 E_3 \over {\sqrt{ (E_2 + E_3)^2 - {\vec p}_{23}^{\, 2}  }
( E_2 + E_3)  }  \right)^{\frac12} ,
\end{eqnarray}
follow from our assumption on normalization of nucleon momentum eigenstates 
$< {\vec p}_i \mid {\vec p}^{\, \prime}_i> = \delta ( {\vec p}_i-{\vec p}^{\, \prime}_i )$ 
 and the action of the permutation operator $P$. 

In a partial wave representation 
the relativistic Faddeev Eq. (\ref{Faddeevrelmom}) 
is explicitly given as \cite{gloeckle86}
\begin{eqnarray}
\phi_\alpha ( p, q)&=&  { 1 \over E_b - {\cal E } (  p ,   q) }  
\sum _ {\alpha ' \alpha '' } 
\int  _ {0 }^\infty d q ' { q '}^2 
\int _ {-1 }^1 d x { T_{\alpha \alpha'} (p,\pi_1; q') \over {{\pi_1}^{l'}}}
\cr & & \times {  G _ { \alpha ' \alpha ''  } ( q , q ', x) \over { 
     {\cal N }_ 1  ( q, q', x ){\cal N }_ 2 ( q, q', x )  }   }  
 { \phi_{\alpha '' } 
 ( \pi_2 ,   {q'} ) \over { \pi_2} ^ {l''}  }  .
\label{Faddeevrelpartial}
\end{eqnarray}
The index $\alpha$ comprises a  set of quantum numbers (channels)
\begin{eqnarray}
\mid \alpha \rangle = 
\mid (ls)j (\lambda {1\over 2 }) I (j I) J (t {1 \over 2})T \rangle , 
\end{eqnarray}
where 
$l$, $s$, $j$ and $ t $ are the orbital angular momentum, 
total  spin, total angular momentum and total 
isospin in the two-body subsystem, respectively. The indices  
$\lambda$, $I$, $J$, and $T$  stand for  the orbital angular momentum,  the 
total angular momentum of the third particle, 
 the total three-body angular momentum,
and the total isospin \cite{gloecklebook}. 
 $G _ { \alpha  \alpha '  } ( q , q ', x) $ results from matrix element of 
the permutation operator and is given by (note that there is a misprint 
 in Eq.~(B2) of ref.~\cite{gloeckle86}) 
\begin{eqnarray}
G_ { \alpha  \alpha '  } ( q , q ', x) = \sum _ {k} 
P_ {k}  ( x)  \sum _ { l_ 1 + l_ 2 = l     }  \, 
 \sum _ { l'_ 1 + l'_ 2 = l'    }  
p^{  l_ 2 +  l'_ 2   } \,  {p '}^{  l_ 1 +  l'_ 1 } \, 
 ( 1 + y_ 1 )^{l_2} \,  ( 1 + y_ 2)^{l'_ 1} \,
g  _ {\alpha \alpha '   }  ^ { {k}  l_ 1   l_ 2   l'_ 1  l'_ 2   } .
\label{permut}
\end{eqnarray}
The expressions for $\pi_1$, $\pi_2$, $y_1$, $y_2$, 
${\cal N }_ 1  ( q, q', x )$ and ${\cal N }_ 2 ( q, q', x ) $
as well as other details can be found  in ~\cite{gloeckle86}.
The matrix elements of the permutation operator $P$ that appear in 
Eqs.(\ref{Faddeevrelpartial}) and (\ref{permut}) correspond to the form of the 
permutation operator given in  \cite{gloecklebook}. There also 
purely geometrical coefficients 
 $g  _ {\alpha \alpha '   }  ^ { {k}  l_ 1   l_ 2   l'_ 1  l'_ 2   }$ are 
derived (see Eqs.~(3.349), (3.352) and (A.19) in that reference). 
 Note that the expressions for the geometrical coefficients 
 are the same in the relativistic and nonrelativistic cases because we neglect
the effect of the Wigner spin rotations.
 This is justified because those effects were found numerically to be 
 of little importance in \cite{witala05}.
Equation (\ref{Faddeevrelpartial}) was then solved in \cite{kamada02} 
with the principal
result that the relativistic binding energies are smaller by 0.3-0.45 MeV
with respect to the nonrelativistic values based on the same 2N potentials.

In many applications the partial wave projected Faddeev components 
$\phi_\alpha (p , q )$ are not 
sufficient and we will show now how to obtain the relativistic 
wave function components $\psi_\alpha ( p , q ) $ from Eq.~(\ref{Psi}).
To this aim we derive and apply a version of the relativistic 
operator $P$ ``working to the right''.
As usual it is sufficient to consider only one overlap, for example
${}_1 \langle {\vec p} \, {\vec q} \mid {\vec p}^{\ '} 
{\vec q}^{\ '} \rangle_2$.
We restrict ourselves to the 3N c.m. system and
express the individual 
 momenta ${\vec p}_{1}$, ${\vec p}_{2}$ and ${\vec p}_{3}$
in terms of ${\vec p}$ and $ {\vec q}$ using Eqs.~(\ref{p2}) and (\ref{p3})
\begin{eqnarray}
{\vec p}_1 =  {\vec q} 
\label{p1.2}
\end{eqnarray}
\begin{eqnarray}
{\vec p}_2 =  {\vec p} - \frac12{\vec q} \, 
+ \frac{{\vec p} \cdot {\vec p}}
{ \left[ 
\omega \left( \vec p \, \right) + \sqrt{ \omega \left( \vec p \, \right)^2 
+ {\vec q}^{\ 2} }
\right]
 \omega \left( \vec p \, \right)  } \, {\vec q} ,
\label{p2.2}
\end{eqnarray}
\begin{eqnarray}
{\vec p}_3 = -{\vec p} - \frac12{\vec q} \, 
- \frac{{\vec p} \cdot {\vec q}}
{ \left[ 
\omega \left( \vec p \, \right) + \sqrt{ \omega \left( \vec p \, \right)^2 
+ {\vec q}^{\ 2} }
\right]
 \omega \left( \vec p \, \right)  } \, {\vec q} ,
\label{p3.2}
\end{eqnarray}
Next we calculate the relative momentum of nucleons 3 and 1  
in their  2N c.m. frame using Eq.~(\ref{p23relative})
\begin{eqnarray}
{\vec p}^{\, \prime\prime} \equiv {\vec \pi} \, ({\vec p}_3, {\vec p}_1) 
\equiv  { 1 \over 2 } \,
\left(  {\vec p}_3  - {\vec p}_1 \right)
- { 1 \over 2 } \, {\vec p}_{31}
\, \left[
 {  E_3  - E_1 \over ( E_3 + E_1 ) +
\sqrt{ (E_3 + E_1)^2 - 
  {\vec p}_{31}^{\ 2} 
  } } \right] ,
\label{p31relative}
\end{eqnarray}
where 
$ {\vec p}_{31} = {\vec p}_{3} + {\vec p}_{1} = -{\vec p}_{2}$.
As a consequence ${}_1 \langle {\vec p} \, {\vec q} \mid {\vec p}^{\ '} {\vec q}^{\ '} \rangle_2$
becomes
\begin{eqnarray}
\label{Pright.1}
{}_1 \langle {\vec p} \, {\vec k} \mid {\vec p}^{\ '} {\vec k}^{\ '} \rangle_2  = 
\left|
\frac{ \partial (  {\vec p}_2 , {\vec p}_3 ) }
     { \partial ( {\vec p} , {\vec p}_{23}  ) }
 \right|^{\frac12} \, 
\left|
\frac
     { \partial ( {\vec p}^{\ ''} , {\vec p}_{31} \,  ) }
     { \partial (  {\vec p}_3 , {\vec p}_1 ) }
\right|^{\frac12} \,
\delta^3 \left( {\vec p}^{\ '} -  {\vec p}^{\ ''} (  {\vec p} , {\vec q} ) \right) \,
\delta^3 \left( {\vec q}^{\ '} -  {\vec p}_2 (  {\vec p} , {\vec q} ) \right) \\ \nonumber
= \, \left( \frac{ 4 E_2 E_3 }{ ( E_2 + E_3) \sqrt{ (E_2 + E_3)^2 - {\vec p}_{23}^{\, 2} }} \right)^\frac12 
\,
\left( \frac{ ( E_3 + E_1) \sqrt{ (E_3 + E_1)^2 - {\vec p}_{31}^{\, 2} }}{ 4 E_3 E_1 } \right)^\frac12 
\\ \nonumber
\delta^3 \left( {\vec p}^{\ '} -  {\vec p}^{\ ''} (  {\vec p} , {\vec q} ) \right) \,
\delta^3 \left( {\vec q}^{\ '} -  {\vec p}_2 (  {\vec p} , {\vec q} ) \right)
\\
\equiv M ( {\vec p}, {\vec q} \, ) \
\delta^3 \left( {\vec p}^{\ '} -  {\vec p}^{\ ''} (  {\vec p} , {\vec q} ) \right) \,
\delta^3 \left( {\vec q}^{\ '} -  {\vec p}_2 (  {\vec p} , {\vec q} ) \right) .
\nonumber
\end{eqnarray}
The scalar function $  M ( {\vec p}, {\vec q} \, ) $ 
actually depends on the magnitudes $ \mid {\vec p} \mid $,  $ \mid {\vec q} \mid $
and the scalar product $ x \equiv  {\hat p} \cdot  {\hat q} $.
Again it is easy to recover the nonrelativistic limit 
of this overlap:
\begin{eqnarray}
 M ( {\vec p}, {\vec q} \, )  \ \rightarrow \ 1
\end{eqnarray}
\begin{eqnarray}
{\vec p}^{\ ''} (  {\vec p} , {\vec q} \, ) \ \rightarrow \ -\frac12 {\vec p} 
- \frac34 {\vec q} ,
\end{eqnarray}
\begin{eqnarray}
{\vec p}_2 (  {\vec p} , {\vec q} \, ) \ \rightarrow \ {\vec p} 
- \frac12 {\vec q} .
\end{eqnarray}
Having obtained Eq.~(\ref{Pright.1}) it is then straightforward to calculate 
the matrix elements of the permutation operator P 
in our standard basis \cite{gloecklebook}
\begin{eqnarray}
\langle \, p \, q \, \alpha \mid P \mid  p' \, q' \, {\alpha}' \, \rangle \ 
= \\ \nonumber
\int _ {-1 }^1 d x \,
\frac{ \delta ( p^{\, '} - \tilde{p} ) }{ \tilde{p}^{\, l^{\, '}+2}} \,
\frac{ \delta ( q^{\, '} - \tilde{q} ) }{ \tilde{q}^{\, {\lambda}^{\, '}+2}} \,
 \tilde{G}_ { \alpha  \alpha '  } ( p , q , x)  \, M ( p, q , x) ,
\end{eqnarray}
where 
\begin{eqnarray}
\tilde{p} \equiv \sqrt{ \frac14 p^2 (1-g)^2 \, + \, \frac9{16} q^2 (1+h)^2 \, + \, \frac34 p q x (1-g) (1+h) } , \\ \nonumber
\tilde{q} \equiv \sqrt{  p^2 \, + \, \frac1{4} q^2 (1+ 2 f)^2 \, - \, p q x (1+ 2 f) },
\end{eqnarray}
\begin{eqnarray}
\tilde{G}_ { \alpha  \alpha '  } ( p , q , x) = \sum _ {k}
P  _ {k}  ( x)  \,
\sum _ { l'_ 1 + l'_ 2 = l'    } \,
\sum _ { \lambda'_ 1 + \lambda'_ 2 = \lambda'    } \,
p^{\,  l'_ 1 + \lambda'_ 1 } \, q^{ \, l'_ 2 + \lambda'_ 2 } \\ \nonumber
 (1 - g)^{l'_ 1} \,  (1 + h)^{l'_ 2} \, 
 ( 1 + 2 f )^{ \lambda'_ 2 } \, \tilde{g}_{\alpha \alpha ' }^{ {k}  l'_ 1   l'_ 2   \lambda'_ 1 \lambda'_ 2   } ,
\end{eqnarray}
\begin{eqnarray}
f \equiv \frac{-p q x }{\left( 2\sqrt{m^2 + p^2} + \sqrt{ 4 m^2 + 4 p^2 + q^2 } \right) 2 \sqrt{m^2 + p^2} } ,
\label{f}
\end{eqnarray}
\begin{eqnarray}
g \equiv {  E_3  - E_1 \over ( E_3 + E_1 ) +
\sqrt{ (E_3 + E_1)^2 - 
  {\vec p}_{31}^{\ 2} 
  } } ,
\label{g}
\end{eqnarray}
and finally
\begin{eqnarray}
h \equiv -\frac23 f \, + \, \frac13 g \, + \, \frac23 f g  .
\label{h}
\end{eqnarray}
The purely geometrical quantity 
$ \tilde{g}_{\alpha \alpha ' }^{ {k}  l'_ 1   l'_ 2   \lambda'_ 1 \lambda'_ 2   } $
is strictly the same (under the neglection of the Wigner spin rotations)
as we use for example in \cite{Pright}.
Consequently, the 3N bound state wave function components can 
be easily calculated.

\begin{figure}[hp]\centering
\epsfig{file=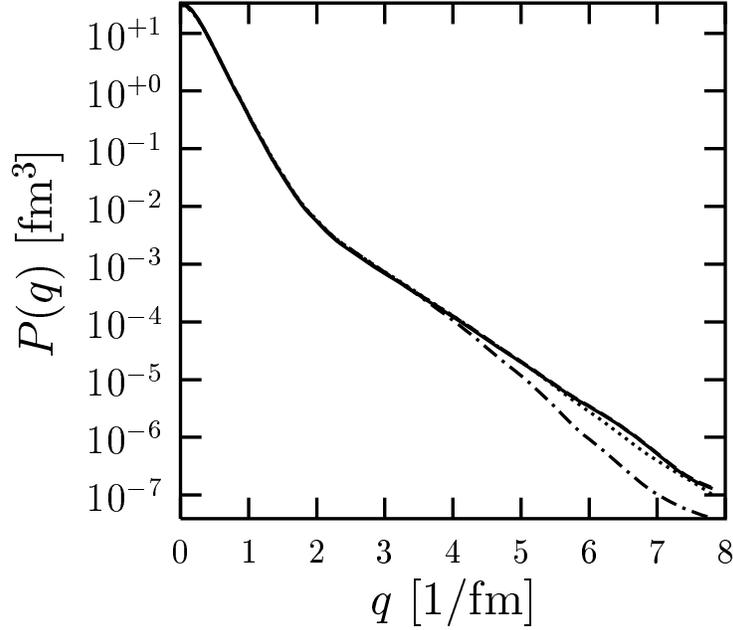,width=10cm}
\caption{
The single-nucleon
momentum distribution for the 3N bound state. 
The curves correspond to strictly nonrelativistic
(dash-dotted), relativistic with no boost effects in the $T$-matrix (dotted),
relativistic with approximate boost effects in the $T$-matrix 
according to Eq.~(\protect\ref{approxboost})
(dashed) and fully relativistic calculations (solid).
}
\label{fig0}
\end{figure}

We would like to give the reader an example of the difference between 
the nonrelativistic and relativistic wave function and show the single-nucleon
momentum distribution in Fig.~\ref{fig0}. 
We see that differences visible on a logarithmic plot
 appear only for 
$ q\gtrsim $ 3 fm$^{-1}$. 
 Most important effects are just due to the relativistic 
kinematics. The approximation given in Eq.~(\protect\ref{approxboost})
does a very good job since the dashed and solid lines nearly overlap. 
 The boost effect is visible for $ q\gtrsim $ 6 fm$^{-1}$. 
The results presented in Fig.~\ref{fig0} and all other results 
in this paper were obtained with the CD Bonn NN potential~\cite{CDBonn}. 
Based on our experience, see for example \cite{golak02},
we expect little sensitivity of our results
to the choice of modern high precision NN potential.

\section{Results for the 
\bm{$ \vec{^3{\rm He}} ({\vec e},e' p)pn$} 
and
\bm{$ \vec{^3{\rm He}} ({\vec e},e' n)pp$} 
processes}
\label{sec:3}

\begin{figure}[hp]\centering
\epsfig{file=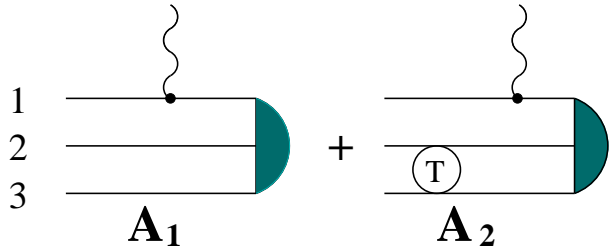,width=8cm}
\caption{
Diagrammatical representation 
of the three-body breakup of $^3$He.
The curly lines denote the photon coupling to nucleon 1.
The large semi-circles depict the initial $^3$He bound state.
While the diagram $A_1$ neglects all the final state interactions 
among the three final nucleons, in the diagram $A_2$
the boosted scattering operator $T$ acts only in the subsystem $(23)$.
}
\label{fig1}
\end{figure}

We will start this section with a brief derivation of the nuclear 
matrix elements
corresponding to Fig.~\ref{fig1}. Here we do not take FSI 
among the three outgoing nucleons fully into account.
In the $A_1$ diagram, which we call the
plane wave impulse approximation (PWIA) in this paper, FSI is 
totally neglected.
In the $A_2$ diagram FSI is restricted only
to one pair of nucleons.
We will denote the approximation
corresponding to the sum of diagrams $A_1$ and $A_2$ by FSI23.
The laboratory frame coincides with the initial 3N c.m. system
so the projection of the relativistic 
wave function on the space of the individual momenta ${\vec p}_i$ 
reads 
\begin{eqnarray}
 \langle {\vec p}_1 \, {\vec p}_2 \, {\vec p}_3 
      \mid \Psi_b \rangle = 
\frac1 {{\cal N }( {\vec p}_2, {\vec p}_3 \, )} \,
\langle \vec\pi ( {\vec p}_2, {\vec p}_3 \, ) , {\vec p}_1 \mid \Psi_b \rangle .
\label{p1p2p3psi}
\end{eqnarray}
Assuming the action of the single nucleon current operator, 
the amplitude $A_1$ takes a very simple form
\begin{equation}
A_1 = \langle {\vec p}_1 m_1 \nu_1 {\vec p}_2 m_2 \nu_2 {\vec p}_3 m_3 \nu_3
      \mid
       j ( {\vec Q} , 1) \mid \Psi_b M M_T \rangle ,
\end{equation}
where ${m_i}$ (${\nu_i}$) 
are spin (isospin) projections of the outgoing nucleons. 
The spin (isospin) magnetic
quantum number of the initial 3N bound state 
is denoted by $M$ ($M_T$). ($M_T = \frac12$
for the $^3$He nucleus.) The single nucleon current
$j ( {\vec Q} , 1)$ acts only on the nucleon 1.
One proceeds by inserting single nucleon intermediate states 
and using (\ref{p1p2p3psi})
\begin{eqnarray}
A_1 = & \delta ( {\vec p}_1 + {\vec p}_2 + {\vec p}_3 - {\vec Q} ) 
\sum_{{m_1}'} j( {\vec p}_1 , {\vec p}_1 - {\vec Q} ; m_1 , {m_1}'; \nu_1)
\nonumber \\
& \langle {\vec p} \, {\vec q} \, {{m_1}'} m_2 m_3 \nu_1 \nu_2 \nu_3
      \mid \Psi_b M M_T \rangle \ \frac1{{\cal N} ( {\vec p}_2 , {\vec p}_3 )} ,
\end{eqnarray}
where $ {\vec p} \equiv {\vec \pi} ( {\vec p}_2 , {\vec p}_3 ) $, 
and   $ {\vec q} \equiv {\vec p}_1 - {\vec Q} $.
Finally we use the partial wave decomposition of the bound state 
in the basis $ \mid p q \alpha \rangle$ and arrive at
\begin{eqnarray}
A_1 = \delta ( {\vec p}_1 + {\vec p}_2 + {\vec p}_3 - {\vec Q} ) \
      \delta_{\nu_1 + \nu_2 + \nu_3 , M_T } \
          \nonumber \\
\ \frac1{N ( {\vec p}_2 , {\vec p}_3 )} \
\sum_{{m_1}'} j( {\vec p}_1 , {\vec p}_1 - {\vec Q} ; m_1 , {m_1}'; \nu_1) 
          \nonumber \\
\sum_{{\alpha}'} \ \sum_{{\mu}'} \ 
     C(j',I',\frac12; {\mu}',M-{\mu}',M) \
     C(l',s',j';{\mu}'-m_2-m_3,m_2+m_3,{\mu}') 
           \nonumber \\
     C(\frac12, \frac12, s'; m_2 , m_3, m_2+m_3) \
     C({\lambda}', \frac12, I' ; M-{\mu}'-{m_1}' , {m_1}' , M-{\mu}' ) 
            \nonumber \\
     C(t',\frac12, \frac12 ; \nu_2 + \nu_3, \nu_1 , \nu_1 + \nu_2 + \nu_3 ) \
     C(\frac12, \frac12, t'; \nu_2 , \nu_3, \nu_2+\nu_3) 
         \nonumber \\
  Y_{ l', {\mu}'-m_2-m_3 } ( {\hat p} ) \
      Y_{ {\lambda}', M-{\mu}'-{m_1}' } ( {\hat q} ) \
       \langle p q {\alpha}' \mid \Psi_b \rangle . \ \ \ \ \ \ \
\end{eqnarray}

The amplitude $A_2$ additionally contains the free 3N propagator $G_0$ 
and the (half-shell) boosted scattering operator $T$ acting in the $(23)$ subsystem
\begin{eqnarray}
A_2 = \langle {\vec p}_1 m_1 \nu_1 {\vec p}_2 m_2 \nu_2 {\vec p}_3 m_3 \nu_3
      \mid T \, G_0 \,
       j ( {\vec Q} , 1) \mid \Psi_b M M_T \rangle 
 \nonumber \\
 = \ \delta ( {\vec p}_1 + {\vec p}_2 + {\vec p}_3 - {\vec Q} ) \
      \delta_{\nu_1 + \nu_2 + \nu_3 , M_T } \
      \delta_{\nu_1 , {\nu_1}'} \ 
\frac1{{\cal N} ( {\vec p}_2 , {\vec p}_3 )} \
         \nonumber \\
\sum_{{m_1}'} j( {\vec p}_1 , {\vec p}_1 - {\vec Q} ; m_1 , {m_1}'; \nu_1) 
          \nonumber \\
 \int \, d {\vec p}^{\ '} \, \sum_{ {m_2}',{m_3}'} \, \sum_{ {\nu_2}',{\nu_3}'} \,
         \delta_{\nu_2 + \nu_3 , {\nu_2}' + {\nu_3}' } \
          \nonumber \\
\langle {\vec p} m_2 m_3 \nu_2 \nu_3 \mid 
T ({\vec p}_2 + {\vec p}_3) 
\mid  {\vec p}^{\ '} {m_2}' {m_3}' {\nu_2}' {\nu_3}' \rangle 
          \nonumber \\
\frac1{ 
       E_2 + E_3 
         \ - \      
         \sqrt{ 4 m^2 + 4 {\vec p}^{\ '\,2} + 
         ({\vec Q} -  {\vec p}_1)^{\ 2} } \ 
 + i \epsilon } \
          \nonumber \\
\langle {\vec p}^{\ '} {\vec q} \, {{m_1}'} {m_2}' {m_3}' {\nu_1}' {\nu_2}' 
{\nu_3}' \mid \Psi_b M M_T \rangle . \ \ \ \ \ \ \ 
\end{eqnarray}
In the final step both the bound state wave function 
and the $T$-matrix are given in the
partial wave basis, which yields 
\begin{eqnarray}
A_2 = \delta ( {\vec p}_1 + {\vec p}_2 + {\vec p}_3 - {\vec Q} ) \
      \delta_{\nu_1 + \nu_2 + \nu_3 , M_T } \
\frac1{{\cal N} ( {\vec p}_2 , {\vec p}_3 )} \
           \nonumber \\
\sum_{{m_1}'} j( {\vec p}_1 , {\vec p}_1 - {\vec Q} ; m_1 , {m_1}'; \nu_1)
           \nonumber \\
\sum_{ l s j \mu t } \
   C(l,s,j;{\mu}-m_2-m_3,m_2+m_3,{\mu}) \ 
   C(\frac12, \frac12, s; m_2 , m_3, m_2+m_3) \
           \nonumber \\
  C(t,\frac12, \frac12 ; \nu_2 + \nu_3, \nu_1 , \nu_1 + \nu_2 + \nu_3 ) \
 C(\frac12, \frac12, t; \nu_2 , \nu_3, \nu_2+\nu_3) \ 
           \nonumber \\
Y_{ l, {\mu}-m_2-m_3 } ( {\hat p} ) \
\sum_{\bar{l}} \, \sum_{{\alpha}'} \, \delta_{l' \bar{l}} \,
               \delta_{s' s} \,
               \delta_{j' j} \,
               \delta_{t' t} \,
     C(j,I',\frac12; {\mu},M-{\mu},M) \nonumber \\
     C({\lambda}', \frac12, I' ; M-{\mu}-{m_1}' , {m_1}' , M-{\mu} ) \
 Y_{ {\lambda}', M-{\mu}-{m_1}' } ( {\hat q} ) 
           \nonumber \\
\int \, d p' \, {p'}^{\ 2} \ 
\langle p (l s )j t \mid T ({\vec Q} -  {\vec p}_1) \mid p' (l' s')j t \rangle \ 
\langle p' q {\alpha}' \mid \Psi_b \rangle 
          \nonumber \\
\frac1{ 
E_2  + E_3 \,
- \, \sqrt{ 4 m^2 + 4 {\vec p}^{\ '\,2} + ({\vec Q} -  {\vec p}_1)^{\ 2} } + i \epsilon } .
\end{eqnarray}

The single nucleon current matrix elements
$ j( {\vec p}_1 , {\vec p}_1^{\ '} ; m_1 , {m_1}'; \nu_1) $
($\nu_1$ decides whether the photon couples 
to a proton or to a neutron)
are taken completely relativistically, i.e.,
\begin{eqnarray}
j( {\vec p} , {\vec p}{\ '} ; m_1 , {m_1}') \equiv
j^\mu ( {\vec p} , {\vec p}^{\ '} ; m_1 , {m_1}') \ =
                 \sqrt{ \frac{m}{ \sqrt{m^2 + p^2 }}} \
                 \sqrt{ \frac{m}{ \sqrt{m^2 + {p}^{\ '2 }}}} 
       \nonumber \\
{\bar u}(p m_1) \left( F_1 \gamma^\mu + i F_2 \sigma^{\mu \nu} 
          (p - p')_\nu \right) u(p' {m_1}') ,
\end{eqnarray}
where $u$ are Dirac spinors. $ F_1 ( p' - p)^2 ) $ and $ F_2 ( p' - p)^2 ) $
are Pauli and Dirac nucleon form factors, respectively.
In this paper we used the H\"ohler parametrization 
for the nucleon electromagnetic form factors~\cite{Hoehler}. 

In this section the results for the three-body breakup will be discussed.
We assume the reference frame for which the
three-momentum transfer $\vec{Q} \equiv {\vec k} - {\vec k}^{\, \prime} $
is parallel to $\hat{z}$,
$\hat{y} \equiv \frac{ {\vec k}^{\, \prime} \times {\vec k}}
{\mid {\vec k}^{\, \prime} \times {\vec k} \mid}$,
and $\hat{x} = \hat{y} \times \hat{z}$.
Here $ {\vec k}$ and ${\vec k}^{\, \prime} $ are the initial and final
electron momenta.
The exclusive cross section for the 
$ e + ^3{\rm He} \rightarrow e^\prime + p + p + n$ reaction has the form~\cite{Donnelly}
\begin{eqnarray}
d \sigma ({\vec S} , h) \ & = & \ 
\sigma_{\rm Mott} \, 
\left\{ \,
\left( v_L W_L + v_T W_T + v_{TT} W_{TT} + v_{TL} W_{TL} \right) \, \right. \nonumber \\
& + & \left. h \, \left( v_{T'} W_{T'} + v_{TL'} W_{TL'} \right) \, 
\right\} \, 
\delta \left( k + m_{^3{\rm He}} - {k}^{\, \prime} - E_1 - E_2 - E_3  \right)  \,
\nonumber \\
&  & 
\delta \left(   {\vec k} - {\vec k}^{\, \prime} -  {\vec p}_1  -  {\vec p}_2  
-  {\vec p}_3  \right)  \,
d^3 {\vec k}^{\, \prime} \, d^3 {\vec p}_1 \, d^3 {\vec p}_2\, d^3 {\vec p}_3 ,
\label{eq4}
\end{eqnarray}
where $\sigma_{\rm Mott}$ and all $v_i$ are analytically given kinematical 
factors,
$h$ is the helicity of the incoming electron
and ${\vec S}$ represents the initial $^3$He spin direction.
The electron mass is neglected and $m_{^3{\rm He}}$ denotes the $^3$He mass. 
The response functions $W_i$,
which contain the whole dynamical information, are constructed from
the nuclear current matrix elements taken between the initial bound state
and the final scattering state. 
Using Eq.~(\ref{eq4}) three observables which we consider in this paper can 
be easily 
constructed. The first one is the unpolarized sixfold differential cross section
\begin{eqnarray}
\frac{d^6\sigma}{d {k}^{\, \prime} \, d{\hat k}^{\, \prime} \, 
dE_1 \, d{\hat p}_1 }
\ = \
\frac12 \, \sum\limits_{m_S} \sum\limits_{m_1, m_2, m_3} \,
{\cal C} \,
\int \!\! d {\hat p} \, {\cal J} \, p_1 E_1 \, \frac14 (E_2 + E_3) 
\, p \, \nonumber \\
\sigma_{\rm Mott} \, 
\left( v_L W_L + v_T W_T + v_{TT} W_{TT} + v_{TL} W_{TL} \right) \,  ,
\label{sigma6}
\end{eqnarray}
where $m_S$, $m_1$, $m_2$, $m_3$ are spin projections of the initial $^3$He 
and of the three outgoing nucleons. The relativistic relative 
momentum ${\vec p} \equiv p {\hat p}$ 
is defined in Eq.~(\ref{p23relative}). The additional factor 
${\cal C}= \frac12$ is necessary 
only if the observed particle is a neutron (the two not detected particles 
are then identical). Note that we changed variables 
according to \cite{gloeckle86}
\begin{eqnarray}
d^3 {\vec p}_1 \, d^3 {\vec p}_2 \, d^3 {\vec p}_3 \ = \ 
{\cal J} \, d^3 {\vec p}_1 \, d^3 {\vec p}_{23} \, d^3 {\vec p} ,
\label{calJ}
\end{eqnarray}
in order to simplify integrations over the
unobserved parameters of the final 3N system.
The kinematical factors in Eq.~(\ref{sigma6}) simplify significantly
in the nonrelativistic limit 
\begin{eqnarray}
   {\vec p} \rightarrow \frac12 \left( {\vec p}_2 - {\vec p}_3 \, \right)  
\nonumber \\
   E_i \rightarrow m \nonumber \\
   {\cal J} \rightarrow 1 
\label{Ei}
\end{eqnarray}

The second and third observables we investigate here are special cases of the 
helicity asymmetry $ A ( {\vec S} ) $
\begin{eqnarray}
A ( {\vec S} \, ) \equiv 
\frac{\sigma ({\vec S} , h=+1) - \sigma ({\vec S} , h=-1)}
{ \sigma ({\vec S} , h=+1) + \sigma ({\vec S} , h=-1) },
\label{A}
\end{eqnarray}
under the same kinematical conditions as the unpolarized 
cross section in Eq.~(\ref{sigma6})
and obtained from the corresponding polarized semi-exclusive 
cross sections  $ \sigma ({\vec S} , h) $.
We consider $A_\parallel$ for $ {\vec S} \parallel  \hat{z}$
and $A_\perp $ for $ {\vec S} \parallel  \hat{x}$.
Further we stick to the so-called parallel kinematics,
for which the finally observed nucleon is ejected parallel to $ \vec{Q} $.
In this case $ W_{TT} = W_{TL} = 0$. This choice of kinematical conditions 
is optimal for the FSI23 approximation. We can expect that under these kinematics,
at least for high energies, the reaction mechanism is dominated 
by the processes 
depicted in Fig.~\ref{fig1}. 

Our nonrelativistic framework \cite{report} allows us to calculate
the initial $^3$He and final scattering states consistently 
using any 3N realistic Hamiltonian and including also many-body current 
operators. There is no such relativistic dynamical framework
available at the moment and in this paper we would like to study 
what are the different effects when some nonrelativistic elements are 
replaced by their relativistic counterparts. 
 We focus on the approximation depicted in Fig.~\ref{fig1} 
and calculate the matrix elements corresponding to diagrams $A_1$ and $A_2$,
first strictly nonrelativistically, secondly using a mixed 
approach \cite{carasco03}
with the nonrelativistic $t$-matrix and wave functions but employing 
relativistic
kinematics and the relativistic 
single nucleon current operator. Finally, we use 
consistently 
the relativistic 3N bound state, kinematics, the boosted $T$-matrix and
the relativistic 
single nucleon current operator, as described in Sec.~\ref{sec:2}. 

\begin{figure}[hp]\centering
\epsfig{file=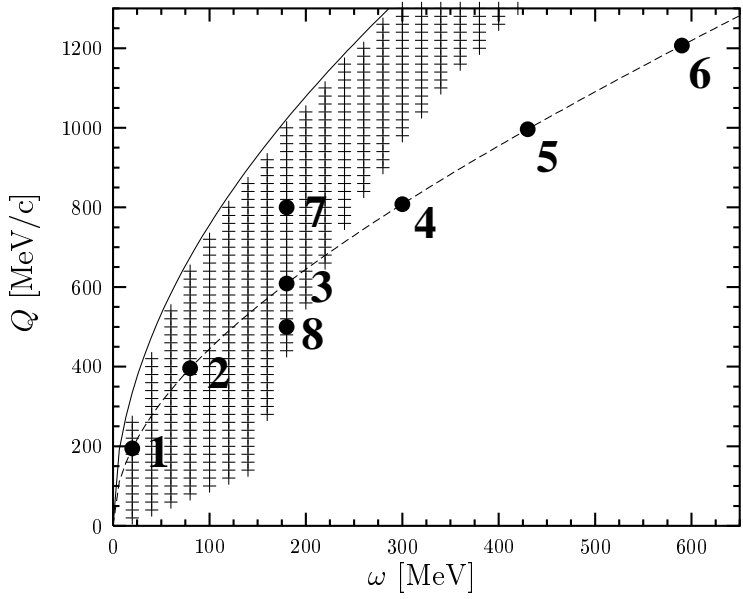,width=10cm}
\caption{
Eight electron kinematics (($\omega$, $Q$) points) considered in the present paper
are marked as full circles.
The shaded area shows the ($\omega$, $Q$) points for which the 3N c.m.
kinetic energy is smaller than the pion mass.
The solid lined corresponds to elastic electron scattering on $^3$He
and the dashed line to the quasi-free scattering condition (scattering on a free nucleon).
}
\label{fig2}
\end{figure}

We chose eight electron kinematics (see Fig.~\ref{fig2} and Table~I), 
characterized by the same electron beam energy ($E$=2000 MeV) and 
different values 
of the energy ($\omega$) and momentum ($Q = \mid \vec Q \mid$) transfers.  
For some of them full inclusion of FSI is possible
within our nonrelativistic framework, since the 3N c.m. energy does not allow 
for pion production. 
In that case we always used the nonrelativistic current operator.
We will thus check to what extent the FSI23 approximation 
might be sufficient
and then concentrate more on different relativistic effects
within this simplified relativistic framework.
For more detailed discussion of the validity 
of the FSI23 approximation see \cite{spectral}. 

\begin{table}[tp]
\begin{center}
\begin{tabular}[t]{ccccccc}
\hline
electron & $\theta_e$ & $E^\prime$ & $\omega$ &   $Q$   & $E_{\rm c.m.}^{\rm 3N} ({\rm rel})$      
&  $E_{\rm c.m.}^{\rm 3N} ({\rm nrl})$         \\[-5pt]
kinematics & [deg]      &   [MeV]      &   [MeV]    &  [MeV/c]  & [MeV]   & [MeV]     \cr
\hline
k1         &  5.6       &   1980   &    20  &  194.8  &   5.6 &    5.5  \cr
k2         & 11.4       &   1920   &    80  &  395.8  &  45.0 &   44.5  \cr
k3         & 17.5       &   1820   &   180  &  608.6  & 109.7 &  106.5  \cr
k4         & 23.5       &   1700   &   300  &  808.3  & 185.4 &  176.3  \cr
k5         & 29.4       &   1570   &   430  &  996.2  & 265.3 &  246.1  \cr
k6         & 36.5       &   1410   &   590  & 1206.7  & 360.9 &  323.8  \cr
k7         & 23.6       &   1820   &   180  &  800.0  &  63.2 &   58.7  \cr
k8         & 14.0       &   1820   &   180  &  500.0  & 130.2 &  127.9  \cr
\end{tabular}
\caption{
Parameters of the eight electron kinematics studied in this paper:
the electron scattering angle $\theta_e$, the outgoing electron energy $E^\prime$,
the energy transfer $\omega$, the magnitude of the three-momentum transfer $Q$,
the relativistic ($E_{\rm c.m.}^{\rm 3N} ({\rm rel})$) 
and nonrelativistic ($E_{\rm c.m.}^{\rm 3N} ({\rm nrl})$) kinetic c.m. 3N energies.
}
\end{center}
\label{tab1}
\end{table}
\begin{figure}[hp]\centering
\epsfig{file=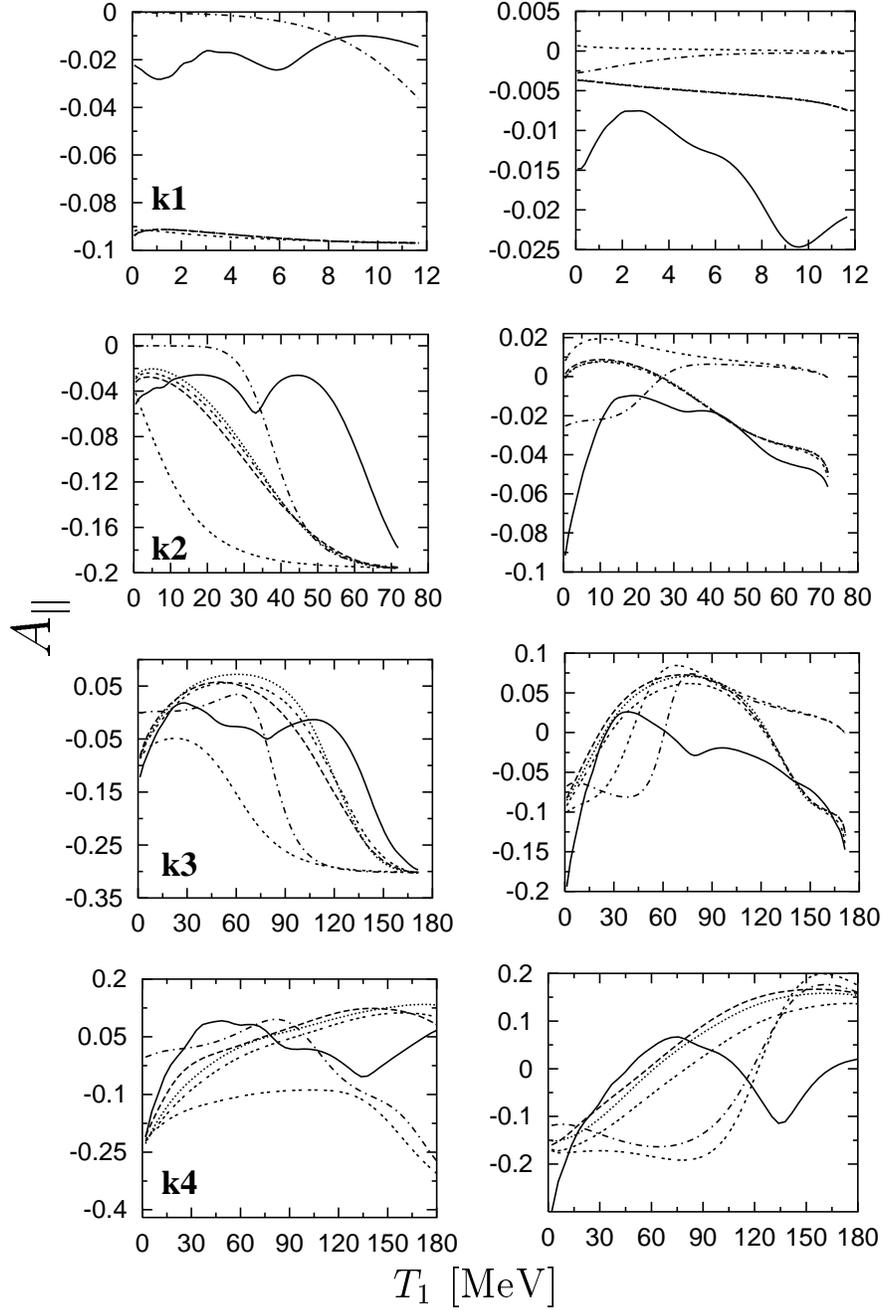,width=12cm}
\caption{
The parallel asymmetry $A_\parallel$ for the neutron (left panel) 
and proton (right panel) ejection in the
virtual photon direction as a function of the emitted nucleon kinetic energy $T_1\equiv E_1 -m$
for the first four electron kinematics from Table~I.
The double dashed line shows the nonrelativistic PWIA prediction 
and the dash-dotted line the nonrelativistic symmetrized PWIA 
(PWIAS) prediction.
Further we show the strictly nonrelativistic FSI23 results (triple dashed line),  
the FSI23 predictions with some relativistic features as described in the text (dotted line),
and the consistent relativistic FSI23 results (dashed line). 
Finally the prediction with full inclusion of FSI is represented by the solid line.
}
\label{fig3}
\end{figure}

\begin{figure}[hp]\centering
\epsfig{file=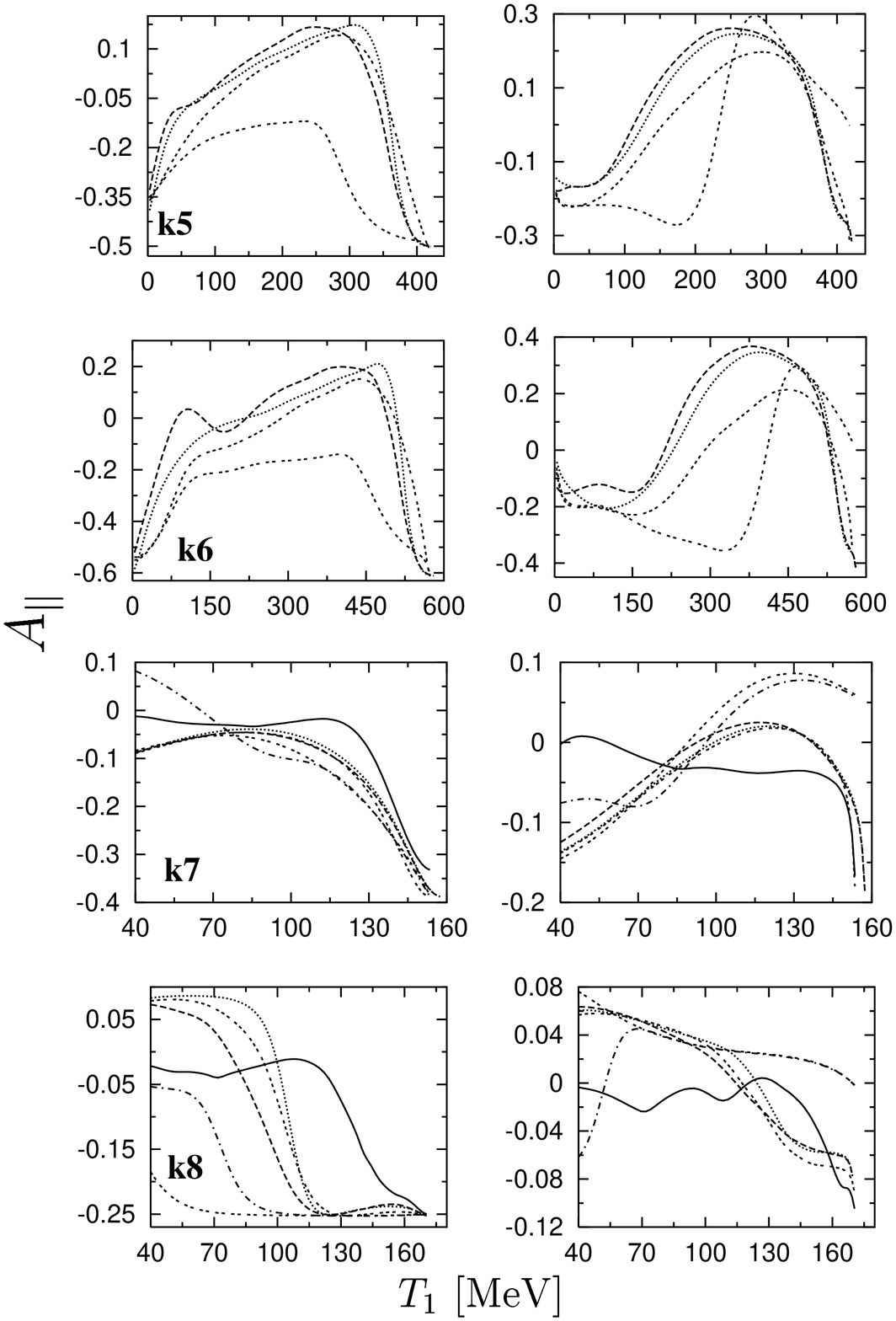,width=14cm}
\caption{
The same as in Fig.~\protect\ref{fig3} for the remaining four electron 
kinematics from Table~I. The PWIAS prediction and the one 
with full FSI are missing for the $k5$ and $k6$ kinematics. 
}
\label{fig4}
\end{figure}

\begin{figure}[hp]\centering
\epsfig{file=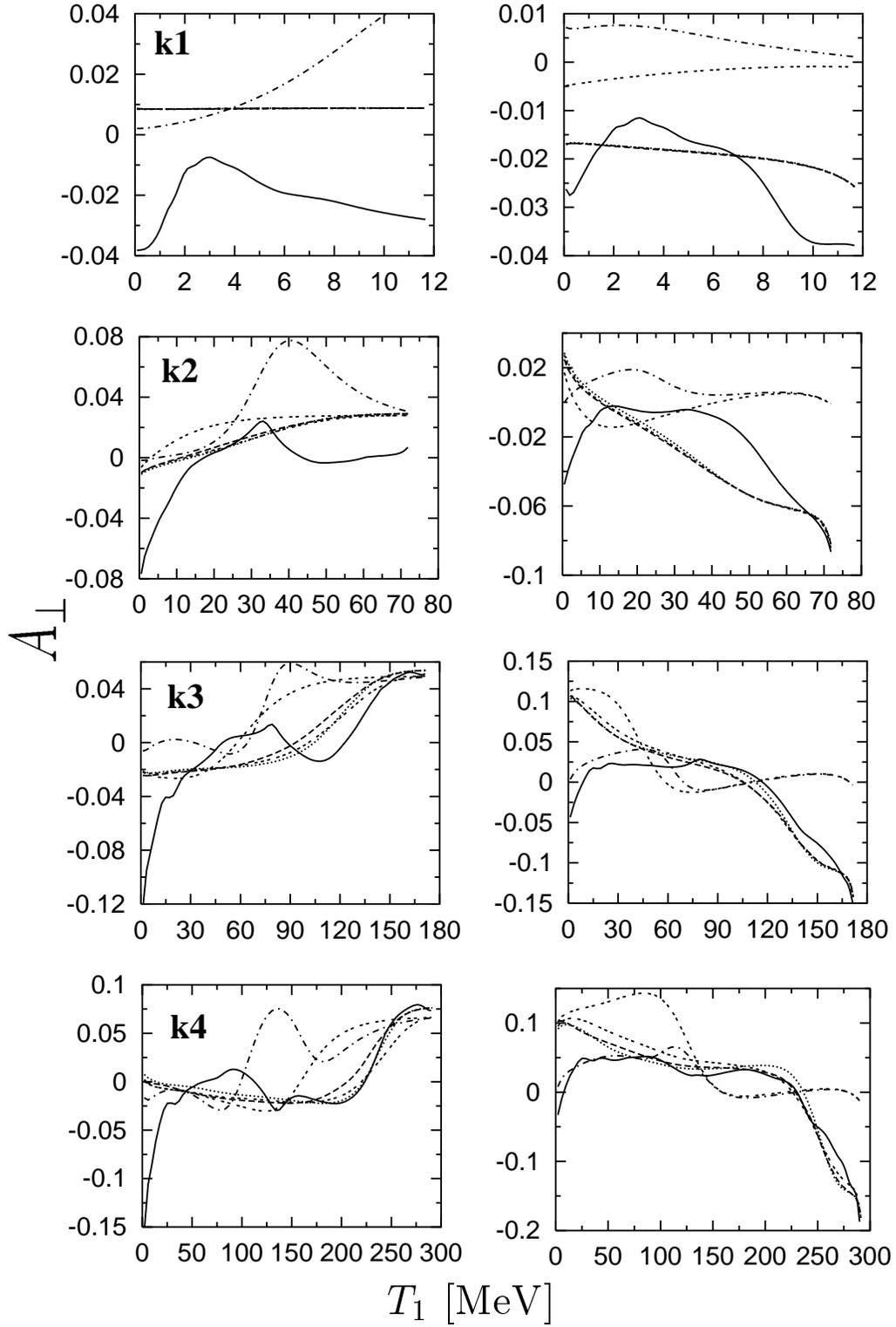,width=14cm}
\caption{
The same as in Fig.~\protect\ref{fig3} 
for the perpendicular asymmetry $A_\perp$.
}
\label{fig5}
\end{figure}

\begin{figure}[hp]\centering
\epsfig{file=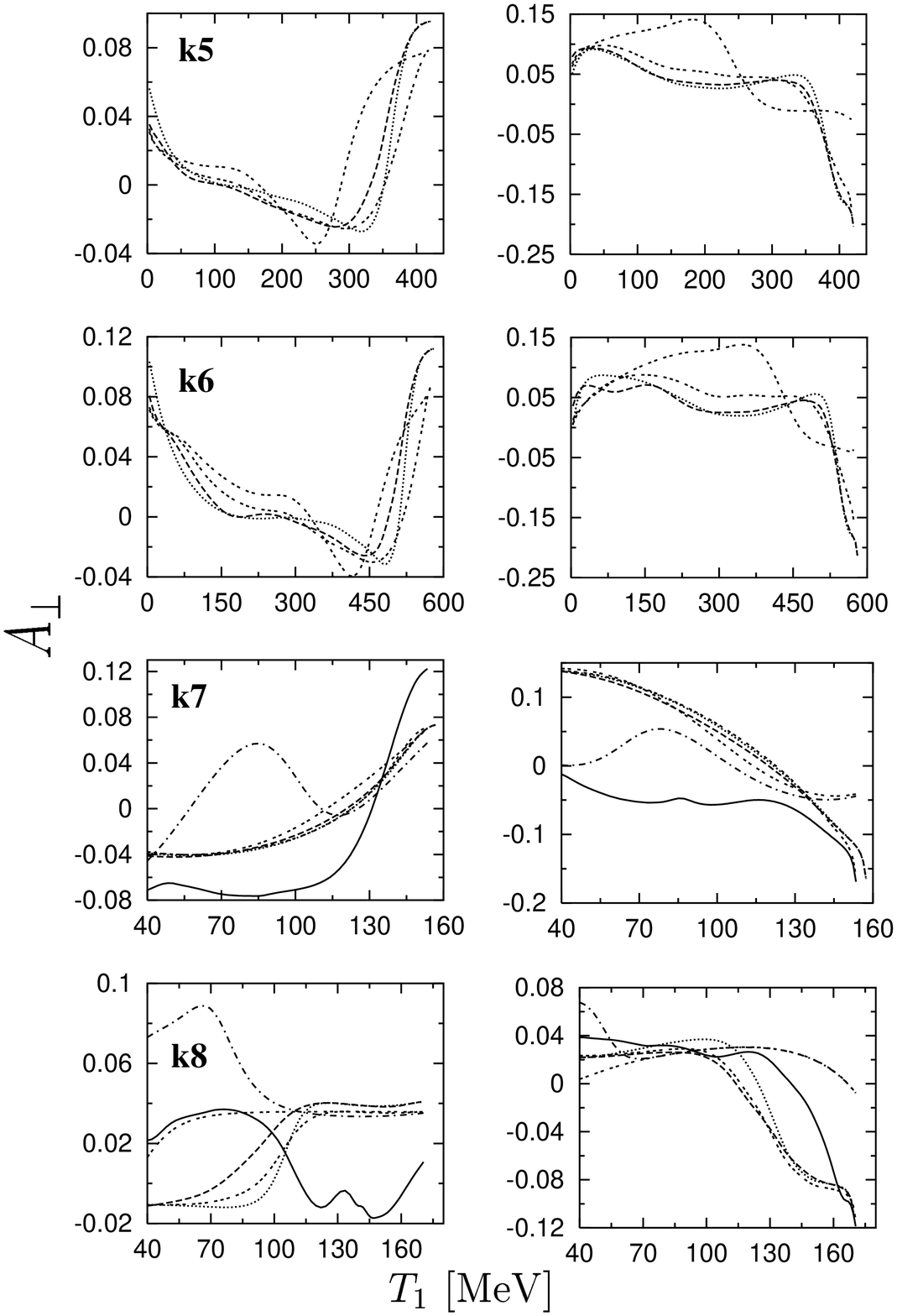,width=14cm}
\caption{
The same as in Fig.~\protect\ref{fig5} for the remaining four electron 
kinematics from Table~I. The PWIAS prediction and the one 
with full FSI are missing for the $k5$ and $k6$ kinematics.
}
\label{fig6}
\end{figure}

\begin{figure}[hp]\centering
\epsfig{file=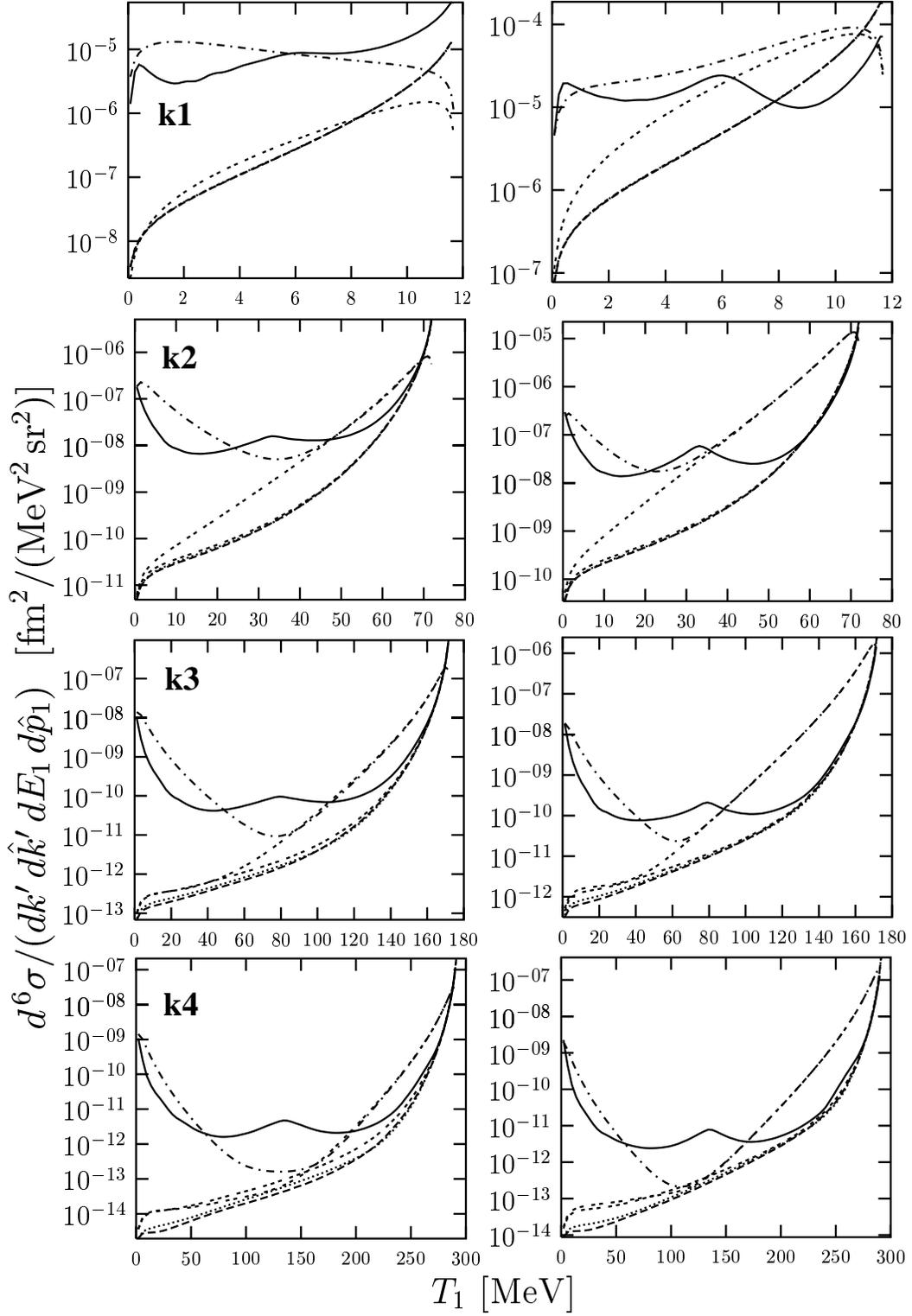,width=14cm}
\caption{
The same as in Fig.~\protect\ref{fig3} 
for the sixfold differential cross section.
}
\label{fig7}
\end{figure}
\begin{figure}[hp]\centering
\epsfig{file=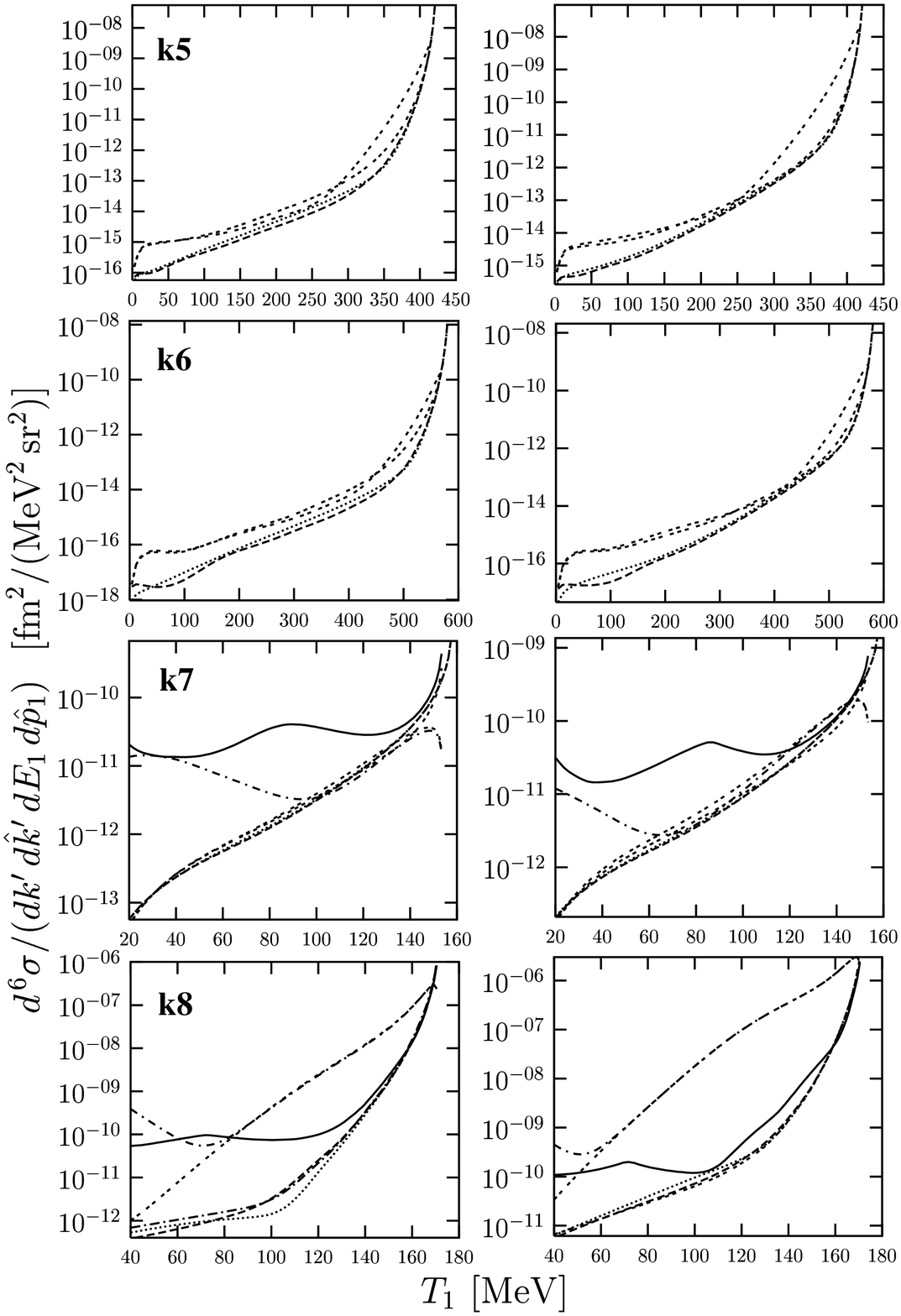,width=14cm}
\caption{
The same as in Fig.~\protect\ref{fig7} for the remaining four electron 
kinematics from Table~I. The PWIAS prediction and the one 
with full FSI are missing for the $k5$ and $k6$ kinematics.
}
\label{fig8}
\end{figure}

The kinematics $k1$-$k6$ are chosen along the quasi-elastic scattering line.
The additional kinematics $k7$ and $k8$ are chosen above 
and below the quasi-elastic line in order to identify differences with respect
to the kinematics that belong to the quasi-elastic scattering group. 

In the first two figures (\ref{fig3} and \ref{fig4}) we show 
the parallel helicity asymmetry 
$A_\parallel$ both for the neutron and proton knockout.
In most cases the FSI23 approximation is not sufficient, 
i.e. the nonrelativistic FSI23 curve lies far away from the 
nonrelativistic prediction taking FSI fully into account.
The latter reveals very often much a more complicated behavior
contrary to the rather simple shapes of the FSI23 predictions.
This FSI23 approximation turns out to be satisfactory (however not always) 
only at the upper end of the energy spectrum for higher magnitudes 
of the three-momentum transfers. 
It is interesting to notice that the contribution 
from the $A_2$ diagram is very small for the $k_1$ kinematics
in the neutron case. Here the PWIA and all FSI23 curves overlap.
That does not mean, however, that all FSI is negligible in this case.
Already the symmetrization in the plane wave predictions 
changes the picture significantly and the results with full inclusion
of FSI are still very different.

The relativistic effects (the spread among the three FSI23 
predictions) are generally most evident not for the maximal energy 
of the ejected nucleon, where the  $(23)$ subsystem c.m. energy is very small, 
but rather in the middle of the nucleon energy range. 
Generally, the mixed approach to the FSI23 calculation is closer 
to the relativistic result than its fully nonrelativistic partner.
Especially for the $k7$ kinematics the difference between the relativistically 
and nonrelativistically calculated maximal energy of the knocked 
out nucleon is clearly visible. 
For the neutron knockout at the $k3$ and $k4$ kinematics 
the asymmetries tend to reach specific values which depend only on the neutron
magnetic form factors and trivial kinematic factors. 
This corresponds very closely to electron scattering
on a free, fully polarized neutron at rest
and was suggested as a way to access the important 
neutron property, since there is no free neutron target in nature.
Note the big differences between the results for the $k3$, $k7$ 
and $k8$ kinematics which all belong to the same energy transfer 
$\omega$ but have different magnitude $Q$ of the three momentum transfer.

These differences are even more true for the 
perpendicular helicity asymmetry  $A_\perp$
displayed in Figs.~\ref{fig5} and \ref{fig6}. 
For this observable, especially in the case of the neutron knockout, 
FSI23 predictions come close to the results fully employing FSI 
only for the $k3$ kinematics, which lies on the quasi-elastic 
scattering curve. 
The FSI23 predictions lie lower ($k7$) or higher ($k8$) 
than the results based on the more complete dynamical model.
Also for this asymmetry the PWIA and FSI23 predictions
take very simple shapes at the first two kinematics, while 
the full inclusion of FSI leads to more complicated structures. 
The perpendicular asymmetry for the neutron
knockout process is very sensitive to the neutron electric form 
factor so also in this case the values for the maximal 
neutron energies, especially at the $k3$ and $k4$ kinematics,
are determined predominantly by the neutron electric form       
factor values. This explains why in the neutron case the parallel 
asymmetry is much bigger than the perpendicular one.

For the $k5$ and $k6$ kinematics there is a clear gap between 
the pure nonrelativistic FSI23 result 
and the predictions employing relativistic kinematics and the 
relativistic current operator. This is partly due to the arguments 
of the electromagnetic form factors, which differ for these two approaches.
In the nonrelativistic case we simply take $ \omega^2 - {\vec Q}^{\, 2}$
which does not correspond to the true four-momentum transfer felt
by the nucleon. In the relativistic case we (exactly) account for the four momentum 
transferred to the nucleon using the following form
\begin{equation}
\left( \sqrt{m^2 + \left( {\vec p} +  {\vec Q} \right)^{\, 2}}  - \sqrt{m^2 + {\vec p}^{\ 2}} \right)^2
-  {\vec Q}^{\ 2} ,
\end{equation}
where $ {\vec p} $ is the nucleon momentum prior to photon absorption.

Finally, in Figs.~\ref{fig7} and \ref{fig8},
we show the six fold differential cross sections
for the neutron and proton knockout. 
Both, for the proton and neutron knockout, the cross sections 
vary by many orders of magnitude. There is always a very steep rise
when the nucleon energy approaches its maximal value
but the cross section for the proton case is always approximately 
factor 10 larger than the corresponding neutron observable.
The PWIA and FSI23 predictions for small $T_1$ values are 
negligible and differ very much both from the PWIAS 
and results taking FSI fully into account.
Except for the $k1$ kinematics, we can always find an energy interval
(at least on the logarithmic scale) where the group of the FSI23 
lines is very close to the curve obtained with the full inclusion 
of FSI (when applicable). As expected, for the $k5$ and $k6$ 
kinematics the difference between
the fully nonrelativistic and the other FSI23 results is best visible. 
The effects which we see for the $k7$ kinematics are magnified 
by the trivial differences in the allowed energy ranges.

\clearpage

\section{Summary}
\label{sec:4}

Many important observables in electron induced breakup of $^3$He 
are measured in kinematical regions, where a nonrelativistic approach is 
 not applicable. Thus  an approximation is needed with some relativistic 
 features included, which can serve as a practical tool 
to analyze results of such experiments. 
One possibility is to extend the so-called FSI23 
approximation (see Sec.~\ref{sec:3})
which has been used since many years to include some relativistic 
ingredients.

In this study we added to this approach a 
consistent relativistic treatment
of the initial 3N bound state, the relativistic single nucleon current 
operator, the relativistic boosted NN scattering operator
and relativistic kinematics.
We studied a number of electron kinematics, mostly on a quasi-elastic 
scattering line, in order to estimate 
the effects of these new relativistic ingredients. We found out that the bulk 
of relativistic effects comes from the relativistic kinematics. Further 
(consistent with kinematics) relativistic features of the calculation 
are less important. 
For the kinematics within the shaded area of Fig.~3 ($k1$, $k2$, $k3$, $k7$, $k8$),
even in the neighborhood of the highest energy of the ejected nucleon,
the FSI23s and the full FSI are different. Therefore the full FSI treatment
is mandatory for a quantitative analysis. On top, especially for the $k7$
and $k8$ kinematics some relativistic effects are noticeable and should be included 
in the future analyzes of correspondingly precise data.
For the kinematics $k5$ and $k6$ in the range of highest nucleon energies
the relativistic effects are clearly visible, especially for neutron emission,
and should be taken into account in the analysis of experimental data.
Whether full FSI effects will be present there, too, cannot be answered by us 
right now. Nevertheless, as the first step
the constructed approximate framework
can be used to analyze experimental data taken
at high energy and momentum transfers.
 In the future  it should be replaced by fully relativistic calculations 
not available at the moment which include all FSI's along the lines of the 
approach applied in the 3N continuum in \cite{witala05,skibinski06}.

\acknowledgments
This work was supported by the Polish Committee for Scientific Research
under grant no. 2P03B00825.
One of us (W.G.) would like to thank the Foundation for Polish Science
for the financial support during his stay in Krak\'ow.
The numerical calculations have been performed
on the IBM Regatta p690+ of the NIC in J\"ulich, Germany.



\begin{thebibliography}{99}

\bibitem{nogga03}  A. Nogga, A. Kievsky, H. Kamada, W. Gl\"ockle, 
                   L.E. Marcucci, S. Rosati, M. Viviani,
                   Phys. Rev. C{\bf 67}, 034004 (2003) and references therein.

\bibitem{report96} W.~Gl\"ockle, H.~Wita{\l}a, D.~H\"uber, H.~Kamada,J.~Golak, 
 Phys. Rep. {\bf 274}, 107 (1996).

\bibitem{kuros02} H.~Wita{\l}a  {\em et al.},  
Phys. Rev. C{\bf 63}, 024007 (2001); 
J.~Kuro\'s-\.Zo{\l}nierczuk {\em et al.}, 
Phys. Rev. C{\bf 66}, 024003 (2002).

\bibitem{report} J.~Golak, R.~Skibi\'nski, H.~Wita{\l}a, W.~Gl\"ockle, 
                 A.~Nogga, H.~Kamada, Phys. Rep. {\bf 415}, 89 (2005).

\bibitem{deltuva04} A. Deltuva, L.P. Yuan, J. Adam Jr., P.U. Sauer, 
                    Phys. Rev. C{\bf 70}, 034004 (2004).

\bibitem{xu00} W. Xu  {\em et al.}, Phys. Rev. Lett. {\bf 85}, 2900 (2000).

\bibitem{extraction} J. Golak, G. Ziemer, H. Kamada, H. 
Wita{\l}a, W. Gl\"ockle,
                  Phys. Rev. C{\bf 63}, 034006 (2001);
                 J. Golak, W. Gl\"ockle, H. Kamada, 
H. Wita{\l}a, R. Skibi\'nski, A. Nogga,
                 Phys. Rev. C{\bf 66}, 024008 (2002). 

\bibitem{sensitivity} J. Golak, W. Gl\"ockle, H. Kamada, 
H. Wita{\l}a, R. Skibi\'nski, A. Nogga,
                      Phys. Rev. C{\bf 65}, 044002 (2002).

\bibitem{xu03} W. Xu {\em et al.},  Phys. Rev. C{\bf 67}, 012201 (2003).

\bibitem{correlations} W. Gl\"ockle, H. Kamada, J. Golak, A. Nogga,
H. Wita{\l}a, R. Skibi\'nski,  and J. Kuro\'s-\.Zo{\l}nier\-czuk,
Acta Phys. Pol. {\bf B32}, 3053 (2001).

\bibitem{spindep} J. Golak, W. Gl\"ockle, H. Kamada, H. Wita{\l}a, 
                  R. Skibi\'nski, A. Nogga,
                  Phys. Rev. C{\bf 65}, 064004 (2002).

\bibitem{polarizations} J. Golak, R. Skibi\'nski, H. Wita{\l}a, W. Gl\"ockle, A. Nogga, and H. Kamada,
                        Phys. Rev. C{\bf 72}, 054005 (2005).

\bibitem{carasco03} C. Carasco {\em et al.}, Phys. Lett. {\bf B559}, 41 (2003).

\bibitem{kamada02} H. Kamada, W. Gl\"ockle, J. Golak, and 
Ch. Elster, Phys. Rev. C{\bf 66}, 044010 (2002).

\bibitem{gloeckle86} W. Gl\"ockle, T.-S. H. Lee and F. 
Coester, Phys. Rev. C{\bf 33}, 709 (1986).

\bibitem{gloeckle86.2} F. Coester, Helv. Phys. Acta {\bf 38}, 7 (1965);
L. M\"uller, Nucl. Phys. {\bf A360}, 331 (1981);
W. Gl\"ockle, L. M\"uller, Phys. Rev. C{\bf 23}, 1183 (1981).

\bibitem{carlson93} J. Carlson, V. R. Pandharipande, R. Schiavilla, Phys. Rev. C{\bf 47},
484 (1993); J. L. Forest, V. R. Pandharipande, J. L. Friar,
Phys. Rev. C{\bf 52 }, 568 (1995); J. L. Forest, V. R. Pandharipande,
J. Carlson, R. Schiavilla, Phys. Rev. C{\bf 52}, 576 (1995).

\bibitem{kamada98} H. Kamada, W. Gl\"ockle, Phys. Rev. Lett. {\bf 80}, 2547 (1998).

\bibitem{allen00} T. W. Allen, G. L. Payne, and Wayne N. Polyzou, Phys. Rev. C{\bf 62}, 054002 (2000).

\bibitem{witala05} H. Wita{\l}a, J. Golak, W. Gl\"ockle, H. Kamada,
                   Phys. Rev. C{\bf 71}, 054001 (2005);
                   H. Wita{\l}a, J. Golak, R. Skibi\'nski, 
   Phys. Lett. {\bf B 634}, 374 (2006).


\bibitem{wallace01} S. J. Wallace, Phys. Rev. Lett. {\bf 87}, 180401 (2001).

\bibitem{keister06} B. D. Keister and W. N. Polyzou, Phys. Rev. C{\bf 73}, 014005 (2006).

\bibitem{gloecklebook} W. Gl\"ockle, {\em The Quantum Mechanical
                       Few-Body Problem} (Springer-Verlag, Berlin, 1983).

\bibitem{Pright} D. H\"uber, H. Kamada, H.~Wita{\l}a, W.~Gl\"ockle,
                 Few-Body Systems {\bf 16}, 127 (1994).

\bibitem{CDBonn} R.~Machleidt, F. Sammarruca, and Y. Song,
                 Phys. Rev. C{\bf 53}, R1483 (1996).

\bibitem{golak02} J.~Golak, W.~Gl\"ockle, H. Kamada, 
H.~Wita{\l}a, R.~Skibi\'nski, A.~Nogga, 
                  Phys.Rev. C{\bf 66}, 024008 (2002).

\bibitem{Hoehler} G. H\"ohler, E. Pietarinen, I. Sabba-Stefanescu, F. Borkowski,
                  G. G. Simon, V. H. Walther, and R. D. Wendling,
                  Nucl. Phys. {\bf  B114}, 505 (1976).

\bibitem{Donnelly} T. W. Donnelly, A. S. Raskin, Ann. Phys. (N.Y.) {\bf 169}, 247 (1986).

\bibitem{spectral} J. Golak, H. Wita{\l}a, R. Skibi\'nski, W. Gl\"ockle, A. Nogga, H. Kamada,
                   Phys. Rev. C{\bf 70}, 034005 (2004).

\bibitem{skibinski06} R. Skibi\'nski, H. Wita{\l}a, J. Golak, nucl-th/0604033.

\end{thebibliography}
\end{document}